\documentclass[pdflatex,sn-mathphys-num]{sn-jnl}

\usepackage{graphicx}
\usepackage{multirow}
\usepackage{amsmath,amssymb,amsfonts}
\usepackage{amsthm}
\usepackage{mathrsfs}
\usepackage[title]{appendix}
\usepackage{xcolor}
\usepackage{textcomp}
\usepackage{manyfoot}
\usepackage{booktabs}
\usepackage{algorithm}
\usepackage{algorithmicx}
\usepackage{algpseudocode}
\usepackage{listings}
\usepackage{longtable}
\usepackage{array}
\usepackage{float}
\usepackage{url}

\begin{document}

\title[AI-Enabled Accounting Information Systems and Fraud Detection]
{Artificial Intelligence-Enabled Accounting Information Systems and Fraud Detection in Nigeria’s Financial Services Sector: The Moderating Role of Natural Language Processing}

\author*[1]{\fnm{Timothy Oluwapelumi} \sur{Adeyemi}}
\email{atimothy.dev@gmail.com}

\affil*[1]{\orgdiv{WeAreGenius Research Institute},
\orgname{Founder and Principal Researcher},
\orgaddress{\city{Lagos}, \country{Nigeria}}}

\author[2]{\fnm{Abigail Omotola} \sur{Ojogbede}}
\email{abigailojogbede@gmail.com}

\affil[2]{\orgdiv{Park University},
\orgname{MBA},
\orgaddress{\country{United States}}}


\abstract{

The rapid digitalisation of financial systems has improved operational efficiency and financial inclusion while simultaneously increasing exposure to sophisticated forms of cyber-enabled fraud and electronic financial misconduct. Conventional auditing systems, which largely depend on retrospective verification and rule-based monitoring, increasingly struggle to address the complexity and speed of modern financial crime. Consequently, financial institutions are progressively adopting Artificial Intelligence (AI)-enabled Accounting Information Systems (AIS) and Natural Language Processing (NLP) technologies to strengthen fraud detection, continuous auditing, and institutional monitoring.

This study examined the influence of AI-enabled AIS on auditing and fraud detection effectiveness within Nigeria’s financial services sector while additionally evaluating the moderating role of NLP. Anchored on the Fraud Diamond Theory and the Technology Acceptance Model, the study adopted a quantitative cross-sectional survey design. Primary data were collected from 186 professionals across banking, insurance, and FinTech institutions in Nigeria. Data were analysed using descriptive statistics, multiple regression, and hierarchical moderated regression techniques.

The findings revealed that AI-enabled AIS significantly improves auditing and fraud detection effectiveness, particularly through prevention, detection, data analysis, and investigative capabilities. The results further indicated that NLP positively moderates the relationship between AI-enabled AIS and auditing effectiveness by improving semantic interpretation and analytical explainability.

The study concludes that AI-enabled AIS and NLP are increasingly important for strengthening fraud governance, regulatory accountability, and institutional trust within emerging digital financial environments.
}

\keywords{
Artificial Intelligence, Accounting Information Systems, Fraud Detection, Natural Language Processing, Financial Governance, Nigeria.
}

\maketitle

\section{Introduction}
\label{sec:introduction}

The accelerating digital transformation of global financial systems has fundamentally reshaped the operational architecture of banking, accounting, auditing, and institutional governance. Over the past decade, advances in cloud computing, electronic payment infrastructures, mobile banking technologies, real-time settlement systems, and financial technology (FinTech) platforms have significantly improved operational efficiency, financial accessibility, and transactional connectivity across modern financial ecosystems. These technological developments have expanded financial inclusion, strengthened digital commerce, and increased the speed and scale of financial intermediation. However, the same innovations have simultaneously intensified institutional exposure to increasingly sophisticated forms of cyber-enabled financial crime and digitally coordinated fraud.

Contemporary financial fraud has evolved far beyond traditional accounting manipulation and isolated transactional irregularities. Modern fraud schemes increasingly involve synthetic identity fraud, phishing operations, account takeovers, insider-assisted misconduct, transaction laundering, social engineering attacks, and coordinated digital deception networks capable of bypassing conventional monitoring structures. The increasing sophistication, velocity, and interconnected nature of these fraud mechanisms have emerged as major governance concerns for financial institutions, regulators, and policymakers globally. Beyond immediate financial losses, digital financial fraud undermines institutional credibility, weakens investor confidence, disrupts regulatory stability, and erodes public trust in digitally mediated financial systems.

The growing complexity of financial crime is particularly significant within emerging economies where rapid technological expansion frequently outpaces institutional governance adaptation and cybersecurity preparedness. Within Sub-Saharan Africa, Nigeria represents one of the most important contexts for examining these developments due to the rapid expansion of its digital financial ecosystem. In recent years, the widespread adoption of smartphones, mobile banking applications, electronic transfer systems, and FinTech platforms has transformed financial service delivery across the country. Regulatory reforms and financial inclusion initiatives have accelerated the migration toward cashless transactions and electronically integrated financial operations, positioning digital financial infrastructure at the centre of Nigeria’s contemporary banking environment.

However, this rapid technological transformation has simultaneously generated new vulnerabilities associated with cyber-assisted fraud, identity manipulation, unauthorised electronic transfers, layered account exploitation, and coordinated digital financial misconduct. Reports issued by financial oversight and regulatory institutions continue to indicate increasing losses associated with electronic fraud and cyber-enabled financial crime across Nigeria’s banking and FinTech sectors. Fraud schemes involving SIM-swap attacks, phishing campaigns, synthetic identities, disclosure manipulation, and unauthorised transactional activities increasingly exploit weaknesses in fragmented monitoring systems and conventional auditing frameworks. Consequently, concerns regarding institutional resilience, fraud governance effectiveness, and the adequacy of existing control mechanisms have intensified across the financial sector.

The growing sophistication of digital financial crime has exposed significant limitations within many traditional auditing and fraud detection systems. Conventional auditing approaches largely depend on retrospective verification procedures, rule-based monitoring structures, periodic sampling techniques, and manually intensive investigative processes. Although these mechanisms remain useful within relatively stable operational environments, they are often inadequate for detecting adaptive fraud patterns characterised by behavioural complexity, high transaction velocity, and continuously evolving manipulation strategies. Static control architectures frequently struggle to identify concealed transactional relationships, behavioural irregularities, and emerging fraud typologies operating across interconnected digital platforms.

As financial transactions become increasingly data-intensive and technologically interconnected, financial institutions require intelligent analytical systems capable of processing large volumes of heterogeneous information in real time while simultaneously identifying suspicious behavioural patterns and operational anomalies. In response to these challenges, Artificial Intelligence (AI) has emerged as an increasingly important component of modern auditing and financial governance systems. AI-enabled Accounting Information Systems (AIS) integrate machine learning algorithms, predictive analytics, automated monitoring architectures, anomaly detection mechanisms, and intelligent decision-support capabilities to strengthen fraud management and institutional oversight \cite{bao2022,chang2022}. Unlike conventional rule-driven systems, AI-assisted auditing architectures support continuous behavioural modelling, adaptive risk assessment, predictive surveillance, and dynamic analytical processing across complex financial environments.

These capabilities substantially improve institutional capacity to detect concealed irregularities, suspicious transactional relationships, and evolving fraud typologies that conventional auditing systems may fail to identify \cite{han2020,aslam2022}. AI integration additionally contributes to operational efficiency by reducing manual investigative workloads, improving monitoring responsiveness, strengthening analytical consistency, and supporting more informed governance decision-making processes. Within digitally intensive financial ecosystems, AI-assisted systems increasingly function not merely as operational tools, but as strategic governance mechanisms supporting institutional transparency, accountability, regulatory compliance, and financial resilience.

Despite growing scholarly attention devoted to AI-enabled auditing systems, the contribution of Natural Language Processing (NLP) to fraud governance remains comparatively underexplored, particularly within emerging financial environments. NLP refers to the capability of computational systems to process, interpret, and extract meaningful insights from unstructured textual information such as audit reports, compliance documentation, customer complaints, transaction narratives, disclosure statements, regulatory filings, and internal organisational communications \cite{pourhabibi2020,craja2020}. This capability is especially important because substantial volumes of organisational intelligence exist in textual rather than purely numerical form.

Within Nigeria’s financial sector, fraudulent activities frequently involve deceptive communication patterns, semantic ambiguity, falsified documentation, disclosure inconsistencies, and socially engineered manipulation that quantitative transactional analysis alone may not adequately detect. NLP-assisted systems strengthen fraud governance by identifying contextual irregularities, abnormal linguistic structures, behavioural indicators, and semantic inconsistencies associated with fraudulent intent. By integrating semantic interpretation with predictive analytics, NLP enhances analytical precision, strengthens investigative capability, and improves the interpretability of AI-generated outputs.

Importantly, NLP also contributes to explainability within intelligent auditing systems. Many advanced machine learning models operate as opaque analytical structures whose outputs may be difficult for auditors, regulators, and institutional stakeholders to interpret. Limited explainability can weaken institutional confidence in AI-assisted decision-making processes and create concerns regarding accountability, transparency, and regulatory defensibility. NLP-enhanced explainability mechanisms help address these limitations by translating algorithmic outputs into comprehensible audit narratives and contextual explanations, thereby strengthening institutional trust, governance accountability, and regulatory transparency within AI-driven financial environments.

Although empirical research relating to AI-enabled fraud detection has expanded considerably in recent years, existing evidence remains heavily concentrated within technologically advanced Western and East Asian economies \cite{qatawneh2021,chang2022}. These institutional environments differ substantially from emerging African financial systems in terms of technological maturity, cybersecurity infrastructure, regulatory integration, and operational governance conditions. Consequently, the applicability of existing findings to Nigeria’s rapidly evolving digital financial ecosystem remains uncertain.

More importantly, limited empirical attention has been devoted to understanding how AI-enabled AIS and NLP capabilities jointly influence auditing effectiveness and fraud governance within emerging financial environments. Existing studies frequently examine AI adoption, intelligent auditing systems, and NLP-assisted analytics independently, while comparatively little evidence exists regarding the moderating role of NLP in strengthening AI-assisted fraud detection capability, interpretability, explainability, and institutional governance effectiveness within African financial systems.

The practical urgency of this issue has become increasingly significant as financial institutions, regulators, and policymakers confront growing concerns surrounding digital banking stability, cybersecurity resilience, anti-fraud governance, and institutional trust within electronically interconnected financial ecosystems. As digital financial transactions continue to expand across Nigeria’s banking and FinTech sectors, the ability of intelligent auditing systems to support proactive fraud prevention, regulatory accountability, and operational transparency has become strategically important for sustaining confidence in the country’s evolving digital financial infrastructure.

This study therefore addresses these gaps by examining the influence of AI-enabled Accounting Information Systems on auditing and fraud detection effectiveness within Nigeria’s financial services sector while additionally evaluating the moderating role of Natural Language Processing. The study contributes to the literature by extending empirical understanding of intelligent auditing systems within emerging economies and by providing insight into how NLP capabilities strengthen fraud monitoring, analytical interpretation, explainability, and governance effectiveness within AI-assisted financial environments.

Accordingly, the objectives of this study are to:

\begin{itemize}
    \item examine the influence of AI-enabled Accounting Information Systems on auditing and fraud detection effectiveness within Nigeria’s financial services sector; and
    
    \item evaluate the moderating role of Natural Language Processing in the relationship between AI-enabled AIS and auditing effectiveness.
\end{itemize}

Accordingly, the study tests the following hypotheses:

\begin{itemize}
    \item \textbf{H$_1$:} AI-enabled AIS exerts a statistically significant positive influence on auditing and fraud detection in Nigerian financial institutions.
    
    \item \textbf{H$_2$:} NLP significantly moderates the relationship between AI-enabled AIS and auditing and fraud detection.
\end{itemize}

\section{Literature Review and Theoretical Framework}
\label{sec:literature}

\subsection{Theoretical Framework}
\label{subsec:theoretical_framework}

This study is anchored on the Fraud Diamond Theory and the Technology Acceptance Model (TAM), which collectively provide a multidimensional theoretical foundation for examining intelligent auditing systems, fraud governance, and technology adoption within digitally evolving financial environments. The integration of these frameworks enables the study to explain both the structural conditions facilitating fraudulent behaviour and the institutional dynamics influencing the adoption and operational effectiveness of AI-enabled auditing technologies within Nigeria’s financial sector.

The Fraud Diamond Theory developed by \cite{wolfe2004} extends the traditional Fraud Triangle by introducing \textit{capability} as a fourth explanatory dimension alongside pressure, opportunity, and rationalisation. The theory argues that fraudulent behaviour is more likely to occur where individuals possess not only financial motivation and opportunity, but also the technical competence, strategic intelligence, and operational capacity required to exploit organisational vulnerabilities successfully. Although originally conceptualised within conventional organisational settings, the theory has become increasingly relevant within digitally interconnected financial ecosystems characterised by electronic transactions, cyber-enabled operations, and complex technological infrastructures.

Within Nigeria’s rapidly expanding digital financial environment, opportunities for fraud increasingly emerge from fragmented monitoring systems, inadequate transactional visibility, cybersecurity weaknesses, inconsistent compliance structures, and uneven technological capacity across institutions. The rapid growth of mobile banking platforms, electronic payment infrastructures, and FinTech services has significantly improved financial accessibility while simultaneously expanding channels through which cyber-assisted fraud and technologically coordinated misconduct may occur. As digital transactions become increasingly complex and transaction volumes continue to expand, conventional auditing systems frequently struggle to provide real-time surveillance capable of identifying adaptive behavioural irregularities and concealed transactional anomalies.

AI-enabled auditing systems address these limitations by strengthening continuous monitoring, predictive analytics, intelligent anomaly detection, automated compliance verification, and dynamic behavioural modelling processes. Unlike traditional auditing mechanisms that rely heavily on static rules and retrospective assessment, AI-assisted systems are capable of processing large volumes of transactional information adaptively and identifying irregularities that may otherwise remain undetected. These capabilities reduce exploitable governance gaps and strengthen institutional responsiveness to emerging fraud risks within digitally intensive financial environments.

Natural Language Processing further extends these capabilities by enabling the analysis of unstructured textual information embedded within audit reports, compliance documentation, disclosure statements, transaction narratives, internal correspondence, customer complaints, and regulatory filings. Whereas conventional auditing systems rely predominantly on structured numerical data, NLP-assisted architectures can identify deceptive linguistic patterns, semantic inconsistencies, abnormal disclosure structures, and contextual irregularities associated with fraudulent behaviour \cite{craja2020,bao2022}. Consequently, NLP expands fraud detection capability beyond transactional analytics into behavioural, interpretive, and communication-oriented dimensions of organisational risk.

An additional challenge associated with advanced AI systems in financial governance relates to algorithmic opacity and limited interpretability. Many machine learning models operate as highly complex analytical systems whose decision-making processes may be difficult for auditors, regulators, and institutional stakeholders to interpret comprehensively. Limited explainability can weaken confidence in automated auditing systems and create concerns regarding transparency, accountability, and regulatory defensibility. NLP-enhanced explainability mechanisms help address these concerns by translating complex algorithmic outputs into comprehensible audit narratives and contextual explanations that improve interpretability and institutional trust.

Complementing the Fraud Diamond Theory, the Technology Acceptance Model developed by \cite{davis1989} explains the institutional factors influencing the adoption and operational use of intelligent auditing technologies. TAM proposes that perceived usefulness and perceived ease of use represent the primary determinants influencing technological acceptance within organisational environments. In the context of Nigeria’s financial sector, the increasing deployment of AI-enabled auditing systems reflects growing institutional recognition that conventional auditing approaches are increasingly inadequate for managing the scale, speed, and complexity of modern digital financial activities.

The perceived usefulness of AI-assisted auditing systems extends beyond operational automation toward broader governance outcomes including improved fraud detection, risk management, compliance coordination, audit transparency, and institutional accountability. NLP capabilities further strengthen this perceived usefulness by improving semantic interpretation, contextual analysis, and explainability within AI-assisted governance systems \cite{handoko2021,ionescu2020}. Institutions are therefore more likely to adopt intelligent auditing technologies where such systems demonstrably enhance monitoring effectiveness, strengthen governance performance, and improve decision-making reliability.

Taken together, the Fraud Diamond Theory and TAM provide a robust conceptual framework for understanding how AI-enabled AIS and NLP capabilities strengthen fraud governance, institutional monitoring, technological adoption, and auditing effectiveness within Nigeria’s evolving digital financial ecosystem.

\subsection{AI-Enabled Accounting Information Systems and Financial Governance}
\label{subsec:ai_ais}

Accounting Information Systems (AIS) have traditionally functioned as organisational mechanisms for transaction processing, financial reporting, record management, and internal control coordination \cite{alomoush2019,chang2022}. However, advances in artificial intelligence technologies have significantly transformed the functional scope of AIS beyond routine accounting operations toward more intelligent forms of auditing, compliance management, predictive analytics, and fraud governance.

The emergence of AI-enabled AIS reflects a broader institutional transition from reactive auditing architectures toward proactive and continuously adaptive governance systems capable of responding to increasingly dynamic financial environments. Machine learning algorithms, intelligent monitoring frameworks, predictive risk models, and automated anomaly detection mechanisms now constitute critical components of modern auditing infrastructures within digitally integrated financial systems.

Within Nigeria’s financial sector, the rapid expansion of electronic banking services, mobile payment platforms, FinTech operations, and real-time transaction systems has significantly increased both transactional complexity and institutional exposure to cyber-enabled financial misconduct. Financial institutions increasingly process vast volumes of transactional data across interconnected digital channels, thereby creating substantial demand for analytical systems capable of identifying suspicious behavioural patterns accurately and efficiently.

Consequently, many financial institutions have increased investment in AI-assisted monitoring technologies to strengthen fraud management, operational oversight, and institutional governance. Intelligent auditing systems are increasingly utilised for anti-money laundering surveillance, predictive transaction analysis, customer risk profiling, compliance screening, and behavioural anomaly detection across digitally intensive financial environments.

Existing empirical studies indicate that AI-enabled auditing systems improve fraud detection effectiveness through automated data integration, predictive analytics, real-time surveillance, and adaptive behavioural modelling \cite{han2020,alam2022}. Intelligent analytical systems aggregate information from multiple transactional sources including mobile banking applications, payment gateways, electronic transfer platforms, digital wallets, and interbank settlement systems. This integration substantially improves transactional visibility and enables more comprehensive identification of irregular financial activities across complex operational environments.

Importantly, the contribution of AI-enabled AIS extends beyond operational efficiency alone. Intelligent auditing systems increasingly support broader governance objectives including institutional transparency, regulatory accountability, fraud prevention, compliance coordination, and organisational resilience. These capabilities are particularly important within Nigeria’s financial environment, where cyber-enabled fraud, interoperability challenges, fragmented digital infrastructures, and evolving regulatory conditions continue to generate substantial governance concerns.

AI-assisted systems additionally strengthen preventive fraud management through automated compliance monitoring, intelligent transaction surveillance, adaptive risk-scoring architectures, and predictive anomaly detection capabilities capable of identifying suspicious behaviour before significant financial losses occur. Such capabilities are increasingly necessary within financial ecosystems characterised by synthetic identity manipulation, SIM-swap attacks, coordinated electronic transfers, phishing campaigns, and cross-platform transactional exploitation.

Despite these advantages, the implementation of intelligent auditing systems across Nigeria’s financial sector remains uneven due to infrastructural limitations, data integration challenges, technical skill deficiencies, implementation costs, and evolving regulatory requirements. These institutional disparities continue to influence the effectiveness, scalability, and governance impact of AI-assisted auditing technologies across emerging financial environments.

\subsection{Natural Language Processing, Explainability, and Fraud Governance}
\label{subsec:nlp}

Natural Language Processing has emerged as an increasingly important component of intelligent auditing and fraud governance systems due to its ability to analyse, interpret, and extract meaningful insights from large volumes of unstructured textual information within organisational environments. Unlike conventional auditing approaches that focus primarily on structured numerical records, NLP enables institutions to process contextual, semantic, and behavioural information embedded within textual communication \cite{pourhabibi2020,kang2020}.

Within financial institutions, substantial organisational intelligence exists in textual form, including audit reports, compliance documentation, disclosure statements, whistleblower reports, customer complaints, internal correspondence, transaction narratives, and regulatory communications. These materials frequently contain behavioural signals, semantic inconsistencies, and contextual indicators associated with financial misconduct that may not be fully observable through quantitative transaction analysis alone.

NLP-enabled systems operationalise these capabilities through semantic interpretation, linguistic modelling, discourse analysis, sentiment evaluation, contextual reasoning, and behavioural pattern recognition \cite{craja2020,bao2022}. These technologies strengthen fraud governance by identifying deceptive communication patterns, irregular disclosure structures, semantic ambiguity, and contextual inconsistencies associated with fraudulent intent or governance deterioration.

Within Nigeria’s financial sector, NLP capabilities are particularly relevant due to persistent challenges involving insider collusion, cyber-enabled manipulation, disclosure opacity, inconsistent compliance practices, and technologically coordinated fraud schemes. In many instances, textual communication patterns may provide early indicators of organisational misconduct before suspicious activities become fully evident within transactional records.

Beyond fraud detection, NLP additionally contributes to explainability and institutional trust within AI-assisted auditing systems. One of the major criticisms of advanced machine learning systems relates to their “black-box” nature, whereby algorithmic outputs may be difficult for auditors, regulators, and institutional stakeholders to interpret comprehensively. Limited explainability can reduce institutional confidence in automated governance systems and complicate regulatory accountability.

NLP-enhanced explainability mechanisms help address these concerns by converting complex analytical outputs into more interpretable narratives and contextual explanations that support audit transparency, regulatory communication, and institutional accountability \cite{ionescu2020}. Consequently, NLP contributes not only to analytical sophistication, but also to improved interpretability, transparency, governance legitimacy, and trust within AI-driven auditing environments.

\subsection{Empirical Review and Research Gap}
\label{subsec:research_gap}

Existing empirical studies generally support the argument that AI-assisted auditing technologies improve fraud detection capability, institutional monitoring, and governance effectiveness within contemporary financial systems. Previous research demonstrates that machine learning-based analytical systems frequently outperform conventional rule-based mechanisms in identifying anomalous transactions, suspicious behavioural patterns, and evolving fraud typologies \cite{bao2022,chang2022}.

Similarly, recent studies have highlighted the growing importance of Natural Language Processing within intelligent auditing environments. \cite{munoko2022} observed that linguistic analysis of managerial disclosures and audit narratives can reveal behavioural indicators associated with deceptive reporting practices and financial irregularities. Related studies further suggest that explainable AI mechanisms and semantic analytics improve transparency, interpretability, and confidence in AI-assisted governance systems.

Despite these contributions, important contextual and theoretical gaps remain within the literature. Much of the existing empirical evidence originates from technologically advanced Western and East Asian economies characterised by comparatively mature digital infrastructures, stronger cybersecurity frameworks, and more integrated regulatory systems. Consequently, findings derived from these institutional environments may not fully reflect the operational realities of emerging African financial ecosystems characterised by infrastructural fragmentation, uneven technological maturity, evolving regulatory conditions, and rapidly expanding digital financial architectures.

Nigeria represents a particularly important context for investigation due to the coexistence of rapid financial digitalisation and persistent governance vulnerabilities involving cyber-enabled fraud, identity manipulation, insider-assisted misconduct, disclosure opacity, and fragmented compliance coordination across interconnected financial platforms. Despite the growing relevance of intelligent auditing technologies within Nigeria’s financial sector, empirical evidence regarding the combined influence of AI-enabled AIS and NLP capabilities on auditing effectiveness remains limited.

Furthermore, although previous studies have examined AI adoption, intelligent auditing systems, fraud analytics, and NLP-assisted monitoring independently, comparatively little empirical attention has been devoted to understanding how NLP capabilities strengthen the effectiveness, interpretability, and governance performance of AI-enabled AIS within emerging financial environments.

Accordingly, this study addresses an important empirical and theoretical gap by examining the influence of AI-enabled Accounting Information Systems on auditing and fraud detection effectiveness within Nigeria’s financial sector while additionally evaluating the moderating role of Natural Language Processing. In doing so, the study contributes to ongoing discussions surrounding intelligent auditing systems, AI-assisted governance, explainable financial analytics, digital fraud management, and technology-driven accountability within emerging economies.

\section{Research Methodology}
\label{sec:methodology}

\subsection{Research Design and Philosophical Orientation}

This study is grounded in a positivist research philosophy, which assumes that organisational and technological phenomena can be objectively observed, measured, and analysed using empirical data and statistical techniques \cite{sekaran2016}. This philosophical stance is appropriate for examining quantifiable relationships among Artificial Intelligence (AI)-enabled Accounting Information Systems (AIS), Natural Language Processing (NLP), and auditing and fraud detection effectiveness within structured financial environments.

Consistent with this orientation, the study adopts a quantitative cross-sectional survey design under a hypothetico-deductive approach. This approach facilitates the formulation and empirical testing of theory-driven hypotheses using observable data. The cross-sectional design enables data collection from multiple financial institutions within a defined time period, allowing the assessment of relationships among AI-enabled AIS capabilities, NLP integration, and auditing effectiveness in Nigeria’s evolving digital financial ecosystem.

Although cross-sectional designs do not fully establish causality, they remain widely adopted in Accounting Information Systems and financial technology research due to their suitability for analysing real-world organisational phenomena across diverse institutional contexts \cite{qatawneh2022,almomani2021}. Accordingly, this design is considered appropriate for identifying statistically significant relationships among intelligent auditing variables within Nigeria’s financial sector.

\subsection{Population, Sampling Technique, and Data Collection}

The target population comprises professionals involved in auditing, accounting, compliance, fraud management, and financial risk governance within financial institutions regulated by the Central Bank of Nigeria (CBN). These include internal auditors, forensic accountants, compliance officers, financial controllers, and risk analysts operating within commercial banks, insurance firms, and FinTech organisations.

According to CBN regulatory reports, Nigeria’s formal financial sector includes commercial banks, microfinance institutions, insurance firms, and rapidly expanding FinTech organisations \cite{cbn2023stat}. These institutions collectively employ professionals responsible for fraud monitoring, financial reporting, and compliance enforcement across major economic centres such as Lagos, Abuja, and Port Harcourt.

A stratified purposive sampling technique was adopted. Stratification ensured representation across banking, insurance, and FinTech sectors, while purposive selection ensured inclusion of respondents with direct experience in AI-assisted auditing and fraud detection systems \cite{sekaran2016}.

A sampling frame of 310 professionals was developed using institutional directories and professional records from the Institute of Chartered Accountants of Nigeria (ICAN) and the Chartered Institute of Bankers of Nigeria (CIBN). Data were collected via an electronically distributed structured questionnaire between February and April 2024. After data screening for incomplete and inconsistent responses, 186 valid responses were retained, representing a response rate of 74.8\%.

\subsection{Measurement Instrument, Validity, and Reliability}

Data were collected using a structured questionnaire adapted from validated instruments in AI adoption, Accounting Information Systems, and fraud analytics literature \cite{handoko2021,qatawneh2021,chang2022,munoko2022,kocsis2019}. The instrument was refined to reflect Nigeria’s financial services context.

The questionnaire consisted of 47 items measuring eight constructs: Data Gathering, Data Analysis, Risk Assessment, Detection, Prevention, Investigation, Auditing and Fraud Detection Effectiveness, and NLP Capability. All items were measured using a five-point Likert scale (1 = Strongly Disagree to 5 = Strongly Agree).

Content validity was established through expert review involving academics and industry professionals in AIS and financial technology. Construct validity was assessed using Exploratory Factor Analysis (EFA) on pilot data (n = 38), which was excluded from the final sample. Items with loadings below 0.60 were removed to improve construct validity.

To assess common method bias, Harman’s single-factor test was applied. The first factor accounted for 34.6\% of variance, below the 50\% threshold, indicating that common method bias was not a major concern.

Internal consistency reliability was evaluated using Cronbach’s alpha. All constructs exceeded the recommended threshold of 0.70, indicating acceptable to excellent reliability \cite{george2003}.

\subsection{Ethical Considerations}

This study adhered to established ethical standards for research involving human participants. Participation was voluntary, and respondents were informed of the academic purpose of the study prior to data collection. Informed consent was obtained electronically from all participants.

To ensure confidentiality, no personally identifiable information was collected. All responses were anonymised and analysed in aggregate form to prevent identification of individuals or institutions. Participants were also informed of their right to withdraw from the study at any time without penalty.

The study adhered to principles of confidentiality, transparency, non-maleficence, and responsible data handling throughout the research process.

\subsection{Analytical Strategy}

Data analysis was conducted using IBM SPSS Statistics Version 27 and AMOS Version 27 in four stages.

First, descriptive statistics (mean, standard deviation, frequency distributions) were used to summarise respondent characteristics and examine variable distributions.

Second, diagnostic tests were performed, including normality tests (Shapiro--Wilk and Kolmogorov--Smirnov), multicollinearity assessment using Variance Inflation Factor (VIF), tolerance statistics, and residual diagnostics.

Third, multiple linear regression analysis was employed to examine the influence of AI-enabled AIS dimensions on auditing and fraud detection effectiveness.

Fourth, hierarchical moderated regression analysis was conducted to test the moderating effect of NLP. Following Baron and Kenny \cite{baron1986} and Cohen et al. \cite{cohen2003}, predictors were entered in Block 1, NLP in Block 2, and the interaction term in Block 3 after mean-centering variables to reduce multicollinearity.

All analyses were conducted at a significance level of \(p < 0.05\). Effect sizes were also reported to enhance interpretation of practical significance.

\section{Results and Discussion}
\label{sec:findings}

\subsection{Demographic Profile of Respondents}

The demographic characteristics of the respondents indicate that the study successfully captured perspectives from professionals actively engaged in auditing, compliance management, fraud monitoring, accounting, and financial risk governance within Nigeria’s financial services sector. Table~\ref{tab:demographic} presents the demographic distribution of the respondents.

\begin{table}[h]
\caption{Demographic Profile of Respondents (\(N = 186\))}
\label{tab:demographic}
\centering
\begin{tabular}{llll}
\hline
\textbf{Category} & \textbf{Sub-Category} & \textbf{Frequency} & \textbf{Percentage} \\
\hline
Gender & Male & 118 & 63.4\% \\
& Female & 68 & 36.6\% \\
\hline
Age Group & 25--34 years & 49 & 26.3\% \\
& 35--44 years & 81 & 43.5\% \\
& 45--54 years & 42 & 22.6\% \\
& 55 years and above & 14 & 7.5\% \\
\hline
Qualification & Bachelor's Degree & 47 & 25.3\% \\
& Master's Degree & 96 & 51.6\% \\
& Doctorate Degree & 43 & 23.1\% \\
\hline
Experience & 1--5 years & 19 & 10.2\% \\
& 6--10 years & 88 & 47.3\% \\
& 11--15 years & 56 & 30.1\% \\
& Above 15 years & 23 & 12.4\% \\
\hline
Sector & Commercial Banking & 77 & 41.4\% \\
& Insurance & 43 & 23.1\% \\
& FinTech & 36 & 19.4\% \\
& Other Financial Services & 30 & 16.1\% \\
\hline
\end{tabular}
\end{table}

The findings reveal that a substantial proportion of respondents possessed postgraduate qualifications and considerable professional experience within financial governance environments. This suggests that the data were obtained from participants with operational familiarity regarding intelligent auditing systems, AI-assisted fraud monitoring technologies, and institutional compliance processes. Respondents from commercial banking institutions constituted the largest proportion of the sample, reflecting the central role of banking institutions within Nigeria’s digital financial ecosystem and their heightened exposure to cyber-enabled fraud risks associated with real-time electronic transaction environments.

\subsection{Descriptive Statistics and Preliminary Analysis}

Table~\ref{tab:descriptive} presents the descriptive statistics for the principal study constructs.

\begin{table}[h]
\caption{Descriptive Statistics of Study Constructs}
\label{tab:descriptive}
\centering
\begin{tabular}{llll}
\hline
\textbf{Construct} & \textbf{Mean} & \textbf{Standard Deviation} & \textbf{Interpretation} \\
\hline
Data Gathering & 3.74 & 0.871 & High \\
Data Analysis & 3.96 & 0.806 & High \\
Risk Assessment & 3.88 & 0.824 & High \\
Detection & 3.93 & 0.791 & High \\
Prevention & 3.89 & 0.768 & High \\
Investigation & 3.71 & 0.918 & Moderate-High \\
Auditing and Fraud Detection & 3.95 & 0.733 & High \\
Natural Language Processing & 3.82 & 0.847 & High \\
\hline
\end{tabular}
\end{table}

The descriptive results indicate generally positive institutional perceptions regarding the effectiveness of AI-enabled AIS and NLP capabilities in auditing and fraud management activities. Among the examined dimensions, Data Analysis recorded the highest mean score (\(\bar{x}=3.96\)), suggesting that respondents perceive AI-assisted analytical systems as particularly valuable for processing, interpreting, and monitoring large volumes of transactional information within digitally intensive financial environments.

Detection and Prevention similarly recorded relatively high mean values, reflecting growing institutional reliance on predictive monitoring systems, anomaly detection architectures, and automated compliance mechanisms for strengthening fraud governance. These findings reinforce the increasing transition from retrospective auditing procedures toward continuous and intelligence-driven monitoring systems within contemporary financial institutions.

By contrast, Data Gathering and Investigation recorded comparatively lower mean values. This may reflect infrastructural fragmentation, data integration limitations, interoperability challenges, and operational constraints affecting the implementation of intelligent auditing systems across some financial institutions within Nigeria. The relatively higher standard deviation observed for Investigation additionally suggests variation in institutional investigative capability and uneven technological maturity across organisations.

\subsection{Diagnostic Testing and Multicollinearity Assessment}

Prior to inferential analysis, diagnostic procedures were conducted to evaluate model suitability and ensure statistical stability. Variance Inflation Factor (VIF) values ranged from 1.842 to 3.416, remaining substantially below the conventional threshold value of 10. Similarly, tolerance statistics exceeded the recommended minimum threshold of 0.10, indicating that multicollinearity was not a significant concern within the regression models.

Residual screening and normality assessments further indicated no severe violations of regression assumptions. Although slight deviations from perfect normality were observed within a few constructs, such variations are relatively common in organisational and behavioural survey research involving perceptual data. Consequently, the dataset was considered statistically appropriate for multivariate regression analysis.

\subsection{Hypothesis Testing and Empirical Findings}

Multiple regression analysis was conducted to examine the influence of AI-enabled AIS dimensions on auditing and fraud detection effectiveness. The results are presented in Table~\ref{tab:regression}.

\begin{table}[h]
\caption{Multiple Regression Results}
\label{tab:regression}
\centering
\begin{tabular}{llll}
\hline
\textbf{Variable} & \textbf{\(\beta\)} & \textbf{t-value} & \textbf{Significance} \\
\hline
Data Gathering & 0.149 & 3.284 & 0.001 \\
Data Analysis & 0.117 & 2.461 & 0.015 \\
Risk Assessment & 0.081 & 1.912 & 0.058 \\
Detection & 0.173 & 3.706 & \(<0.001\) \\
Prevention & 0.384 & 8.917 & \(<0.001\) \\
Investigation & 0.096 & 2.118 & 0.036 \\
\hline
\multicolumn{4}{l}{\textbf{Model Statistics:} \(R = 0.791;\ R^2 = 0.626;\ \text{Adj. } R^2 = 0.611;\ F = 49.28;\ p < 0.001\)} \\
\hline
\end{tabular}
\end{table}

The regression model demonstrated substantial explanatory power, with the AI-enabled AIS dimensions collectively accounting for approximately 62.6\% of the variance in auditing and fraud detection effectiveness. Within organisational and behavioural research contexts, this represents a comparatively strong explanatory outcome, indicating that intelligent auditing capabilities contribute meaningfully to institutional fraud governance processes.

Among the predictor variables, Prevention emerged as the strongest determinant of auditing and fraud detection effectiveness (\(\beta = 0.384, p < 0.001\)). This finding suggests that predictive monitoring systems, automated compliance mechanisms, behavioural anomaly detection architectures, and proactive surveillance capabilities play particularly important roles in strengthening fraud governance within digitally interconnected financial environments. The finding further reflects the increasing institutional emphasis on preventive fraud management strategies capable of identifying suspicious behavioural patterns before financial losses materialise.

Detection similarly demonstrated a statistically significant positive effect (\(\beta = 0.173, p < 0.001\)), reinforcing the importance of continuous monitoring systems and intelligent surveillance architectures in identifying irregular transactional behaviour across complex financial platforms. Data Gathering and Data Analysis also produced statistically significant positive relationships, indicating that institutional access to integrated financial information and AI-driven analytical capability substantially improve fraud monitoring effectiveness and audit responsiveness.

Risk Assessment exhibited a positive but marginally insignificant relationship with auditing effectiveness (\(p = 0.058\)). Although the coefficient remained positive, the finding suggests that the effectiveness of AI-assisted risk assessment may depend on complementary institutional conditions such as governance maturity, implementation quality, organisational readiness, data reliability, and regulatory integration. This may further indicate that predictive risk-scoring mechanisms alone are insufficient unless supported by broader institutional monitoring frameworks and adaptive operational controls.

Pearson correlation analysis additionally revealed moderate-to-strong positive associations among the AI-enabled AIS dimensions and auditing effectiveness variables, thereby providing further support for the regression findings and reinforcing the internal consistency of the observed relationships.

\subsection{Moderating Influence of Natural Language Processing}

Hierarchical moderated regression analysis was conducted to evaluate the moderating role of Natural Language Processing within the relationship between AI-enabled AIS and auditing effectiveness.

\begin{table}[h]
\caption{Hierarchical Moderated Regression Results}
\label{tab:moderation}
\centering
\begin{tabular}{llll}
\hline
\textbf{Model} & \textbf{\(R^2\)} & \textbf{\(\Delta R^2\)} & \textbf{Significance} \\
\hline
Model 1: AI \(\rightarrow\) Fraud Detection & 0.584 & 0.584 & \(<0.001\) \\
Model 2: AI + NLP & 0.619 & 0.035 & 0.002 \\
Model 3: AI \(\times\) NLP Interaction & 0.632 & 0.013 & 0.041 \\
\hline
\end{tabular}
\end{table}

The moderation results indicate that Natural Language Processing contributes positively to auditing and fraud detection effectiveness within AI-assisted financial environments. The inclusion of NLP capability improved the explanatory performance of the regression model, while the interaction term produced a modest but statistically significant increase in explanatory power.

These findings suggest that NLP strengthens AI-enabled auditing systems by extending analytical capability beyond structured numerical information into unstructured textual environments such as audit reports, compliance documentation, disclosure narratives, customer complaints, managerial communications, and regulatory correspondence. Through semantic interpretation, contextual analysis, and linguistic modelling, NLP enhances institutional capability to identify behavioural inconsistencies, deceptive communication patterns, and contextual irregularities associated with fraudulent activities.

However, the relatively moderate interaction effect suggests that NLP primarily functions as a complementary analytical capability that enhances, rather than independently determines, the effectiveness of intelligent auditing systems. This finding is theoretically important because it indicates that the effectiveness of NLP is maximised when integrated within broader AI-enabled monitoring architectures rather than deployed as an isolated technological mechanism.

\subsection{Discussion of Findings}

The findings of this study demonstrate that AI-enabled Accounting Information Systems (AIS) significantly contribute to auditing effectiveness, fraud monitoring, and institutional governance within Nigeria’s financial services sector. The regression results revealed that AI-assisted prevention, detection, data analysis, and investigative capabilities exert positive and statistically significant influences on auditing and fraud detection effectiveness. These findings suggest that intelligent auditing systems strengthen institutional capacity to identify suspicious transactional behaviour, monitor operational anomalies, and improve responsiveness to emerging fraud risks within digitally interconnected financial environments.

The results further indicate that preventive capability emerged as the strongest predictor of auditing and fraud detection effectiveness. This finding underscores the increasing importance of proactive fraud governance mechanisms capable of identifying suspicious behavioural patterns before financial losses materialise. The result reflects the broader transformation from retrospective auditing approaches toward predictive and continuous monitoring architectures within modern financial systems. As digital transactions become increasingly complex and high-frequency, financial institutions require adaptive analytical systems capable of detecting evolving fraud typologies in real time.

These findings are consistent with \cite{bao2022}, who observed that AI-assisted analytical systems significantly improve anomaly detection, predictive surveillance, and institutional monitoring effectiveness within digitally intensive financial environments. Similarly, the findings align with the work of \cite{han2020}, which reported that machine learning-driven auditing systems improve fraud detection accuracy through automated behavioural analysis and intelligent anomaly identification. The findings additionally support the arguments of \cite{aslam2022}, who emphasised that AI-enabled monitoring architectures enhance institutional responsiveness and strengthen governance efficiency in technology-driven financial systems.

The study also established that Natural Language Processing (NLP) positively moderates the relationship between AI-enabled AIS and auditing effectiveness. The moderation results suggest that NLP enhances the analytical scope of intelligent auditing systems by extending fraud detection capability beyond structured numerical information into unstructured textual environments such as audit reports, transaction narratives, compliance documentation, customer complaints, and regulatory disclosures. Through semantic interpretation, contextual analysis, and linguistic modelling, NLP improves investigative precision, analytical explainability, and fraud pattern recognition within AI-assisted governance systems.

This finding corroborates the work of  \cite{craja2020}, who argued that NLP-assisted systems improve fraud identification by detecting semantic inconsistencies and deceptive communication patterns embedded within organisational documentation. The result also aligns with  \cite{munoko2022}, who found that linguistic analysis of audit narratives and disclosure statements strengthens the detection of irregular reporting behaviour and governance anomalies. Furthermore, the findings support the position of  \cite{ionescu2020}, who emphasised that NLP contributes significantly to explainability and transparency in AI-driven auditing environments by translating algorithmic outputs into interpretable audit narratives.

From a theoretical perspective, the findings extend the Fraud Diamond Theory by demonstrating that AI-assisted auditing systems reduce opportunities for fraudulent behaviour through continuous monitoring, predictive surveillance, behavioural anomaly detection, and automated compliance verification mechanisms. The results additionally reinforce the Technology Acceptance Model (TAM) by suggesting that financial institutions increasingly perceive intelligent auditing technologies as useful where they improve operational efficiency, strengthen fraud governance capability, and enhance institutional accountability within digitally evolving financial ecosystems.

Overall, the findings position AI-enabled AIS and NLP as increasingly important components of contemporary fraud governance and intelligent auditing systems. Within emerging digital financial environments such as Nigeria, the integration of AI-driven analytical systems and NLP-enhanced interpretability mechanisms appears increasingly essential for strengthening institutional resilience, regulatory accountability, operational transparency, and trust within modern financial governance architectures.

\section{Conclusion, Recommendations, and Future Research}
\label{sec:conclusion}

This study examined the influence of AI-enabled Accounting Information Systems (AIS) on auditing and fraud detection effectiveness within Nigeria’s financial services sector while additionally evaluating the moderating role of Natural Language Processing (NLP). The findings demonstrated that AI-enabled AIS significantly strengthens fraud governance, institutional monitoring, and auditing effectiveness within digitally evolving financial environments. Specifically, prevention, detection, data analysis, and investigative capabilities were found to exert positive and statistically significant effects on fraud detection effectiveness, with preventive capability emerging as the strongest predictor. These findings highlight the growing importance of proactive and continuous monitoring systems capable of identifying suspicious behavioural patterns before financial losses materialise.

The study further established that Natural Language Processing positively moderates the relationship between AI-enabled AIS and auditing effectiveness. By extending analytical capability beyond structured numerical data into unstructured textual environments, NLP improves semantic interpretation, contextual analysis, investigative precision, and audit explainability. This suggests that NLP enhances not only analytical sophistication, but also institutional transparency, regulatory accountability, and confidence in AI-assisted auditing systems.

From a theoretical perspective, the findings reinforce the relevance of the Fraud Diamond Theory and the Technology Acceptance Model within contemporary digital financial ecosystems. The study demonstrates that intelligent auditing technologies reduce opportunities for fraudulent behaviour through predictive analytics, automated monitoring, and anomaly detection mechanisms, while also supporting institutional acceptance of technology-driven governance systems due to their perceived usefulness and operational value.

Based on these findings, the study recommends that financial institutions within Nigeria continue to strengthen investment in AI-enabled auditing infrastructure, integrated fraud analytics platforms, and NLP-assisted monitoring systems. Regulatory agencies should also encourage the development of explainable AI frameworks and standardised governance policies capable of improving transparency, interoperability, and accountability in intelligent financial systems. In addition, continuous professional training in AI-assisted auditing, digital compliance monitoring, and data governance should be prioritised to improve institutional readiness and technological adaptability across the sector.

Despite its contributions, the study was limited by its cross-sectional design and reliance on perceptual survey data obtained from regulated financial institutions. Future studies may therefore adopt longitudinal or mixed-method approaches to examine the long-term organisational implications of intelligent auditing systems. Further research may also explore emerging areas such as explainable AI, blockchain-assisted auditing architectures, federated learning systems, and cybersecurity resilience frameworks within fraud governance environments across emerging economies.


\end{document}